\documentclass[sn-mathphys,Numbered]{sn-jnl}

\usepackage{graphicx}%
\usepackage{multirow}%
\usepackage{amsmath,amssymb,amsfonts}%
\usepackage{amsthm}%
\usepackage{mathrsfs}%
\usepackage[title]{appendix}%
\usepackage{xcolor}%
\usepackage{textcomp}%
\usepackage{manyfoot}%
\usepackage{booktabs}%
\usepackage{algorithm}%
\usepackage{algorithmicx}%
\usepackage{algpseudocode}%
\usepackage{listings}%
\usepackage{subfig}

\usepackage{ dsfont }
\usepackage{fancyhdr}

\begin{document}

\title[TRUSWorthy]{TRUSWorthy: Toward Clinically Applicable Deep Learning for Confident Detection of Prostate Cancer in Micro-Ultrasound}


\author*[1,4]{\fnm{Mohamed} \sur{Harmanani}}\email{mohamed.harmanani@queensu.ca}

\author[1,4]{\fnm{Paul} \sur{F. R. Wilson}}

\author[2,4]{\fnm{Minh} \sur{Nguyen Nhat To}}

\author[1,4]{\fnm{Mahdi} \sur{Gilany}}

\author[1,4]{\fnm{Amoon} \sur{Jamzad}}

\author[2,4]{\fnm{Fahimeh} \sur{Fooladgar}}

\author[3]{\fnm{Brian} \sur{Wodlinger}}

\author[2]{\fnm{Purang} \sur{Abolmaesumi}}
\equalcont{These authors contributed equally to this work.}

\author[1,4]{\fnm{Parvin} \sur{Mousavi}}
\equalcont{These authors contributed equally to this work.}

\affil[1]{\orgname{Queen's University}, \orgaddress{\city{Kingston}, \country{Canada}}}

\affil[2]{\orgname{University of British Columbia}, \orgaddress{\city{Vancouver}, \country{Canada}}}

\affil[3]{\orgname{Exact Imaging}, \orgaddress{\city{Markham}, \country{Canada}}}

\affil[4]{\orgname{Vector Institute},
\orgaddress{\city{Toronto}}, \country{Canada}}


\abstract{
    \textbf{Purpose}: While deep learning methods have shown great promise in improving the effectiveness of prostate cancer (PCa) diagnosis by detecting suspicious lesions from trans-rectal ultrasound (TRUS), they must overcome multiple simultaneous challenges. There is high heterogeneity in tissue appearance, significant class imbalance in favor of benign examples, and scarcity in the number and quality of ground truth annotations available to train models. Failure to address even a single one of these problems can result in unacceptable clinical outcomes. 
    \textbf{Methods}: We propose TRUSWorthy, a carefully designed, tuned, and integrated system for reliable PCa detection. Our pipeline integrates self-supervised learning, multiple-instance learning aggregation using transformers, random-undersampled boosting and ensembling: these address label scarcity, weak labels, class imbalance, and overconfidence, respectively. We train and rigorously evaluate our method using a large, multi-center dataset of micro-ultrasound data.
    \textbf{Results}: Our method outperforms previous state-of-the-art deep learning methods in terms of accuracy and uncertainty calibration, with AUROC and balanced accuracy scores of 79.9\% and 71.5\%, respectively. On the top 20\% of predictions with the highest confidence, we can achieve a balanced accuracy of up to 91\%. \textbf{Conclusion}: The success of TRUSWorthy demonstrates the potential of integrated deep learning solutions to meet clinical needs in a highly challenging deployment setting, and is a significant step towards creating a trustworthy system for computer-assisted PCa diagnosis.
}


\keywords{Uncertainty estimation, deep ensemble, prostate cancer, ultrasound}



\maketitle

\pagestyle{fancy}
\fancyhead[LE]{\textit{IJCARS preprint}}
\fancyhead[LO]{}
\fancyhead[RO]{\textit{TRUSWorthy}}
\fancyhead[RE]{}
\renewcommand{\headrulewidth}{0pt}

\section{Introduction}\label{intro}

Early and accurate diagnosis of prostate cancer (PCa) greatly increases the chances of successful treatment. The current standard for the diagnosis and grading of PCa is the histopathological analysis of tissue retrieved from the prostate during biopsy, typically performed under the guidance of trans-rectal ultrasound (TRUS). Conventional ultrasound has a low sensitivity in identifying cancerous lesions~\cite{ahmed2017diagnostic}, meaning that freehand prostate biopsy is typically \emph{systematic}, where tissues are sampled from predefined anatomical locations with the hope of sampling cancerous tissue if any is present. Conversely, clinical state of the art involves \emph{targeted} biopsy where suspicious lesions are identified with pre-procedure multi-parametric magnetic resonance imaging (mp-MRI) using the PIRADS protocol~\cite{ahmed2017diagnostic}, and directly targeted with TRUS-guided biopsies. This greatly increases sensitivity of PCa diagnosis compared to systematic biopsy~\cite{ahmed2017diagnostic}; however, MRI imaging requires specialized facilities and personnel, inherently restricting its use to large, well-funded urban centers. There is a pressing need to develop biopsy targeting methods that rely solely on ultrasound.

The combination of deep learning (DL) with \emph{enhanced} ultrasound imaging modalities such as Contrast-Enhanced Ultrasound (CeUS)~\cite{halpern2006contrast}, temporal-enhanced ultrasound (TeUS)~\cite{fooladgar2022uncertainty}, multi-parametric ultrasound (mp-US)~\cite{postema2015multiparametric}, elastography~\cite{pallwein2007real}, and micro-ultrasound (micro-US)~\cite{klotz2020can} is a promising avenue to improve ultrasound-based PCa detection. Enhanced ultrasound captures richer tissue information than conventional ultrasound, while DL models can learn features from large volumes of high-dimensional ultrasound data, overcoming the limitations of human visual interpretation. 
Micro-ultrasound~\cite{klotz2020can} is a newer modality that utilizes high imaging frequencies ($\sim$29~MHZ) to achieve significantly higher spatial resolution than conventional ultrasound, improving the visualization of tissue microstructure and enabling the identification of cancerous lesions with sensitivity comparable to PIRADS~\cite{ahmed2017diagnostic}. Numerous studies have established the potential of DL for analyzing CeUS~\cite{feng2018deep}, TeUS~\cite{fooladgar2022uncertainty}, mp-US~\cite{soerensen2021deep}, and $micro$-US~\cite{shao2020improving} for various medical imaging goals, including PCa detection.

Despite their promise, DL models for PCa detection contend with four key challenges: (i) \textit{Weak labeling}: ground-truth histopathology labels describe the overall pathology of an entire tissue sample, providing at best, an approximation of the localization of a tumor within an ultrasound image; (ii) \textit{Label scarcity}: only a small proportion of available ultrasound data have corresponding pathology annotations. These two shortcomings result in a significant lack of large, reliably annotated data required for building DL models.
(iii) \textit{Class imbalance}: the balance of data is highly skewed in favor of benign samples with under-representation of aggressive cancer; (iv) \textit{Data heterogeneity}: caused by highly variable appearance of tissues, clinical acquisition protocols, and patient populations, high heterogeneity means models certain to face very unfamiliar data at test time. While models should ideally respond to such data with uncertainty, they are prone to producing highly overconfident and incorrect outputs~\cite{guo2017calibration}. Together, these challenges significantly limit the robustness, generalizability, and trustworthiness of standard DL approaches.

In the literature, multiple instance learning (MIL) has been proposed as an effective solution to address weak labels where multiple patches from a given image are combined into a ``bag" of examples, and the models learn the association between the ``bag" and the corresponding coarse label. This paradigm is currently dominant in digital histopathology~\cite{li2021multi,bulten2022artificial}, and has been explored in ultrasound~\cite{fujita2022weakly,gilany2023trusformer}. For label scarcity, self-supervision has been consistently gaining traction and has been successfully applied to ultrasound data~\cite{ali2023self}. While the problem of class imbalance has been studied in the context of general machine learning~\cite{seiffert2009rusboost}, research is more limited on formal approaches to address this in PCa detection where most studies use simple majority undersampling~\cite{gilany2023trusformer, fooladgar2022uncertainty}. Finally, to avoid overconfident and incorrect predictions, uncertainty estimation approaches such as CRISP~\cite{judge2022crisp}, cascade networks~\cite{xie2022uncertainty}, and deep ensembles~\cite{lakshminarayanan2017simple,fooladgar2022uncertainty} have been proposed in ultrasound imaging. In summary, while the above mentioned challenges have been studied individually, there is a notable lack of \emph{integrated solutions} that simultaneously tackle all the shortcomings and produce clinically useful and robust PCa detection models.

Our core vision is a computer-assisted system for the analysis of interventional TRUS imaging to identify cancerous lesions. However, the failure to account for even a single one of the aforementioned challenges could lead to unacceptable clinical failures through incorrect and overconfident predictions. To this end, we propose TRUSWorthy, an integrated system which to our knowledge is the first method to simultaneously address label scarcity, weak labeling, class imbalance \emph{and} data heterogeneity in PCa detection using ultrasound. TRUSWorthy incorporates components of self-supervised learning using convolutional networks, multiple instance learning using transformers\footnote{these first two components follow the methodology of our previous work~\cite{gilany2023trusformer}}, random undersampled boosting and deep ensembles, which respectively address the four aforementioned issues, into a synergistic cancer detection methodology. On a multi-center dataset of over 600 patients, our approach outperforms the state-of-the-art (SOTA) in PCa detection from micro-ultrasound: it achieves an AUROC of up to $79.9\%$, and has excellent uncertainty calibration, enabling a clinically practical ``reject" option for uncertain predictions. Through these advances, this work makes a major step towards realizing the vision of trustworthy DL systems in PCa diagnosis.

\section{Materials}\label{mats}

We use private data from 693 patients in five clinical centers who underwent TRUS-guided prostate biopsy as part of a clinical trial (NCT02079025). The PSA for inclusion was below 50, and the clinical stage was below cT3. The median age is 63 and the median PSA is 5.8. The procedures are performed using the ExactVu micro-ultrasound imaging system (Exact Imaging, Markham, Canada). For each biopsy sample, a single radio-frequency (RF) ultrasound image with a depth of $28~mm$ and a width of $46.6~mm$ in the sagittal plane is recorded prior to firing the biopsy gun. 
Based on pathology findings, we assign the label 1 (clinically significant cancer; Gleason Score $\geq 7$) or 0 (Benign). Following previous studies~\cite{shao2020improving,gilany2023trusformer}, cores with a low involvement of cancer (less than 40\% of the biopsy tissue by area based on the full core length) are excluded from this study for each patient. Cores with clinically insignificant cancer (GS6) were acquired but subsequently discarded before the dataset was finalized and handed over for model development.

Table~\ref{tab:dataset_summary} shows a detailed summary of the dataset, such as the number of cores from each clinical center and the breakdown by Gleason scores. 
We stratify patients into training and test sets using a 5-fold cross-validation scheme where each split has a proportionate representation of data from all centers. 85\% of the data is used for training, and the remaining 15\% is used for testing. Each training fold is further divided into training (85\%) and validation (15\%) sets, corresponding to a holdout validation set size of 12\% for each fold. 

To extract ROIs, we first identify the needle trace region for each biopsy core on the corresponding ultrasound image. We then divide the ultrasound image into a grid, and determine the positions of several overlapping ROIs ($5\times 5~mm$) by identifying the intersection of the needle and prostate masks. As done in prior works~\cite{rohrbach2018high, shao2020improving}, we experimented with various ROI sizes such as $3mm$, $5mm$, and $7mm$, and found $5mm$ to be the best choice. We selected ROIs using a sliding window centered along the biopsy track with a stride of $1\times1~mm$ between each ROI. ROIs are considered to be inside the needle if they overlap with the region by at least 66\%~\cite{gilany2023trusformer}, a value determined through hyperparameter tuning, which offered the best compromise between patch quality and training diversity. 55 ROIs are extracted from each core and resized from $1780\times55$ to $256\times256$ pixels by performing upsampling in the lateral direction and downsampling in the axial direction. We then rescale the pixel values of each ROI to the range of $(0,1)$, using the ROI's local statistics. 

\begin{table}[t]
\label{tab:dataset_summary}
\caption{Total summary of the dataset, divided by clinical center.}\label{tab:data}%
\begin{tabular}{@{}llllllll@{}}
\toprule
\textbf{Center} & \textbf{Patients} & \textbf{Cores} & \textbf{Benign} & \textbf{GS7} & \textbf{GS8} & \textbf{GS9} & \textbf{GS10} \\
\midrule
JH    & 60 & 616  & 568 & 32 & 10 & 6 & 0  \\
UVA    & 236   & 2335 & 2018 & 221 & 57 & 28 & 11  \\
PCC & 171 & 1599 & 1400 & 162 & 23 & 14 & 0 \\
PMCC & 71 & 588 & 486 & 90 & 12 & 0 & 0 \\
CRCEO    & 155   & 1469  & 1255 &  170 & 32 & 12 & 0 \\
\midrule
\textbf{Total} & 693 & 6607 & 5727  & 675 & 134 & 60 & 11 \\
\botrule
\end{tabular}
\label{tab:dataset_summary}
\end{table}

\begin{figure}[htb!]
    \centering
    \includegraphics[width=\columnwidth]{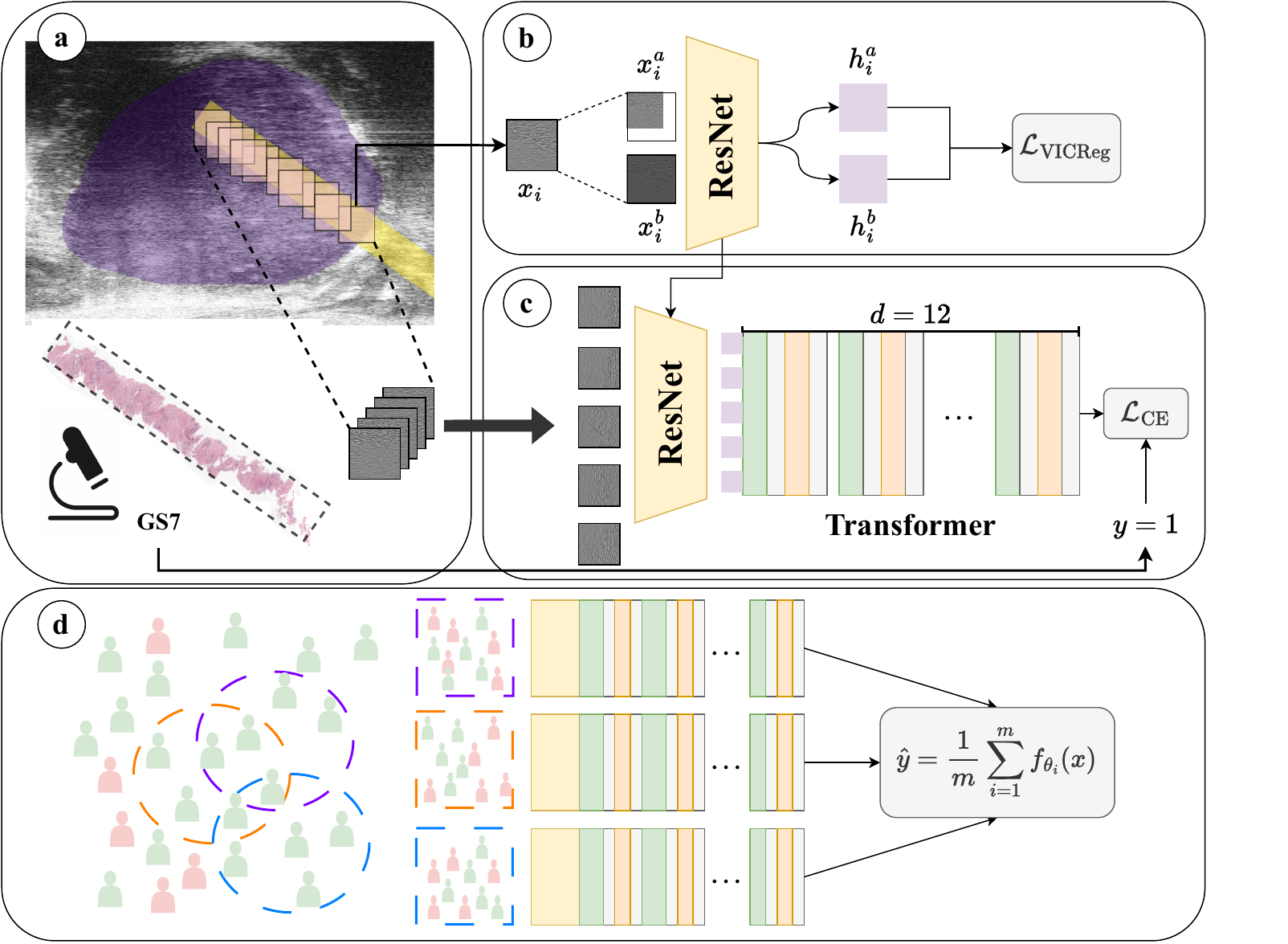}
    \caption{An overview of our proposed approach. \textbf{(a)} Data extraction and coarse labelling from histopathology. \textbf{(b)} Pre-training an ROI classifier using self-supervised learning. \textbf{(c)} MIL finetuning using transfer weights and a Transformer. \textbf{(d)} Training an ensemble of specialized learners on distinctly resampled training sets.}
    \label{fig:workflow}
\end{figure}

\section{Methods}\label{methods}

\subsection{Learning Patch Embeddings from Unlabeled Data}
We use the Variance-Invariance-Covariance Regularization~\cite{bardes2021vicreg} (VICReg) framework for self-supervised learning (SSL) to pre-train a modified ResNet18\footnote{We used 1 convolution per residual block instead of 2.} model on micro-ultrasound data. Each training sample $x_i$ is augmented into two random yet correlated views, $x_i^{a}$ and $x_i^{b}$. We use a combination of standard data augmentations like random crops and flips, as well as augmentations designed specifically for ultrasound such as phase shift and envelope distortion~\cite{wilson2023self}. The ResNet is used to compute embeddings $h_i^a$ and $h_i^b$ for each augmented view, which are then projected onto a latent space ($z_i^j = \text{MLP}(h_i^j)$) by a projection network. The VICReg loss function is then applied~\cite{bardes2021vicreg}.

\subsection{Multiple Instance Learning from Coarse Labels}
After pretraining the feature extractor, we train an MIL aggregator network on the embedding space learned in the self-supervised training stage. For each core, we collect all of the ROIs from the core into a ``bag" of size $n \times 256 \times 256$, where $n$ is the number of ROIs in the core. We extract 512 features from each ROI using the pretrained ResNet model. The resulting bag of features ($55 \times 512$) is then used as the input to a Transformer~\cite{vaswani2017attention} network with 12 layers, 8 attention heads, an inner dimension of 512, and an MLP dimension of 512\footnote{We found those parameters to be the best. We tested depths of 4, 6, 8, and inner dimensions of 128, 256.}. The model's output is a single prediction $\hat{y}$ representing the probability that the core contains cancer. The entire network (feature extractor and aggregator) is now trained end-to-end to optimize the cross-entropy loss between $\hat{y}$ and the corresponding pathology label $y$ for the core.

\subsection{Mixed Deep Ensembles}
The deep ensembles framework trains several models independently from one another, each one with different weight initializations. Given $m$ members with parameterizations $\{\theta_i | i = 1, 2, ... m\}$, the ensemble makes predictions by averaging the members' set of predictions as: $\hat{y} = \frac{1}{m} \sum_{i=0}^m f_{\theta_i} (x)$.

As mentioned in previous sections, the training data suffers from a severe label imbalance, with over 6 benign cores for every malignant core. Experimentally, we observed that training a model with the imbalanced data resulted in degraded performance, and that better performance could be achieved by undersampling the benign set to a fixed benign-to-cancer ratio (we use 2:1 benign to cancer ratio). However, in doing so, we discard a large number of benign cores which are potentially informative training examples. 

To address this issue, we adopt an approach similar to RUSBoost~\cite{seiffert2009rusboost}: for each member of the ensemble, we train the member on a subset of the training data consisting of all of the cancerous cores and a randomly subsampled set of training cores chosen to achieve a 2:1 benign to cancer ratio. In contrast to the conventional ensemble approach wherein identical
models are trained with a different initialization, each member is also trained on a different
subset of benign cores, diversifying and increasing the data points presented to the models. This workflow is shown in Figure~\ref{fig:workflow}d. 
This approach has two benefits: first, the ensemble has been collectively exposed to more benign cores as training examples; secondly, since different members are specialized for different benign examples, the consistency of all members' predictions at inference time is a good measure of predictive confidence. Specifically, as is standard practice~\cite{lakshminarayanan2017simple} we use the \emph{maximum softmax probability} ($MPS = \max(\hat{y}, 1-\hat{y})$), as a measure of confidence. We then apply a threshold to remove the least confident (most uncertain) predictions.

\subsection{Experimental Design}
We integrate each individual method above into one framework (as shown in Fig.~\ref{fig:workflow}), combining their strengths to create a model that addresses several issues that were previously
tackled individually. We chose to use deep ensembles, as it is a proven method for effective uncertainty calibration for our task and dataset~\cite{fooladgar2022uncertainty}. While there exists other more recent methods for uncertainty, deep ensembles pair nicely with RUSBoost, a necessary addition to achieve robustness to data imbalance and scarcity, a problem that has been encountered in PCa detection from ultrasound but never properly addressed~\cite{gilany2022towards, gilany2023trusformer}. Moreover, by exposing the ensemble members to different training sets, RUSBoost increases the diversity of the ensemble members' knowledge, a characteristic that is correlated with better predictive uncertainty~\cite{gontijo2021no}. Finally, we use TRUSformer for our backbone model, as it is the current state-of-the-art (SOTA) method for PCa detection from ultrasound, and solves the problem of noisy labels using MIL~\cite{gilany2023trusformer}.

We train the ResNet feature extractor using VICReg for 200 epochs with a batch size of 64 and the NovoGrad optimizer. We use a learning rate of $10^{-5}$. We tried a combination of standard data augmentations and ultrasound-specific augmentations, but found that the training process was not sensitive to the choice of augmentation. We then finetune the ResNet model on the task of detecting cancer in individual ROIs. We train it for 15 epochs, with the same learning rate and batch size, but using the Adam optimizer and a cosine annealing schedule for the learning rate. We use the fine-tuned ResNet as the backbone for an ensemble of TRUSformer networks, each of them to be trained on a different dataset. We sample 10 different sets of benign cores equal to twice the size of the set of cancer cores and train a TRUSformer network on each set. We train for 75 epochs, using a batch size of 8, a learning rate of $10^{-4}$, and the Adam optimizer. The full range of hyperparameters considered and our hyperparameter tuning approach is outlined in our public code repository, available at: \href{https://github.com/mharmanani/trusworthy}{github.com/mharmanani/trusworthy}.

We compare TRUSWorthy to several baseline methods for PCa detection and uncertainty estimation. As the first baseline, similar to previous work, we train a single ResNet to detect cancer in individual ROIs. We refer to this type of model as an ``ROI-scale" classifier, in contrast to MIL methods which are ``core-scale". For the second ROI-scale baseline, we implement a Deep Ensemble model with 10 members, using the above ResNet as the backbone. We train these models both with and without self-supervision (SSL). 
For the MIL baselines, we compare our method to TRUSformer~\cite{gilany2023trusformer}, the previous SOTA in PCa detection, as well as another Deep Ensemble implementation, using TRUSformer as a backbone. All models are trained and evaluated with a 5-fold cross-validation scheme. 

For evaluating PCa detection, we report metrics such as AUROC, Balanced Accuracy, Sensitivity, and Specificity. For evaluating uncertainty estimation, we report performance at various rejection thresholds, as well as uncertainty calibration metrics such as the Brier Score~\cite{guo2017calibration} and Expected Calibration Error~\cite{guo2017calibration} (ECErr). All metrics are averaged across folds for test data, and the standard deviation is reported. For qualitative model evaluation, we generate heatmaps of model outputs overlaid on corresponding ultrasound images, as follows: First, we utilize a 8$mm$ by 8$mm$ square window sliding in strides of $0.5~mm$ by $0.5~mm$ to generate candidate windows $\mathcal{W} = \{W_1, W_2, ..., W_n\}$ covering the image. We filter the regions not inside the prostate via $\mathcal{W}_\text{filt.} = \{ W \in \mathcal{W} : |W \cap \text{prostate}| / |W| \geq 0.8 \}$. For the $i$'th window we generate a prediction $\hat{y}_i$ by collecting 16 $5mm$ by $5mm$ inside the window and feeding them through our MIL classifier. The uncertainty $u(\hat{y})$ is computed and compared to a threshold $\tau$ to remove uncertain predictions. Finally, the heatmap value $h(x, y)$ at pixel $x,y$ is given by
\begin{equation}
     h(x,y) = \bigg( \sum_{W_i \in \mathcal{W}_\text{filt}} \hat{y}_i \mathds{1}_{\{(x, y) \in W\}} \mathds{1}_{\{u(\hat{y}_i) < \tau\}} \bigg) / \bigg( \sum_{W_i \in \mathcal{W}_\text{filt}} \mathds{1}_{\{(x, y) \in W\}} \mathds{1}_{\{u(\hat{y}_i) < \tau\}} \bigg )
\end{equation}
The opacity of the heatmap $\alpha(x, y)$ at pixel $x, y$ is proportional to the number of confident predictions: 
$\alpha(x,y) \propto \sum_{W_i \in \mathcal{W}_\text{filt}}  \mathds{1}_{\{(x, y) \in W\}} \mathds{1}_{\{u(\hat{y}_i) < \tau\}}.$

\section{Results}\label{sec4}
\textbf{\textit{Cancer Detection: }} Table~\ref{tab:results_perf1} shows the results of our experiments, with the top rows for ROI classifiers and the bottom rows for MIL classifiers. The models' performance is calculated by biopsy core for both ROI-based and MIL methods. We first note the impact of SSL, increasing the AUROC of ResNet by $1.2\%$ and Deep Ensembles by $0.5\%$. SSL is also a key component of the TRUSformer and TRUSWorthy approaches, as the transformer only performs well when trained on VICReg features~\cite{gilany2023trusformer}. We speculate that SSL improves performance by reducing the tendency of the feature extractor to overfit to the limited, weakly labeled training data.

We also observe the benefits of MIL, with an improvement of $2\%$ and $3.3\%$ in AUROC and balanced accuracy when comparing TRUSformer to SSL-ResNet, as well as a major boost when comparing TRUSformer to the baseline ResNet ($+3.2\%$ AUROC, $+1.4\%$ B.Acc.). We believe the improved performance is a result of the models' capacity to learn from coarse and noisy pathology labels.

The addition of ensembles yields further benefits, with TRUSformer-Ens. enjoying an improvement in AUROC of $0.9\%$ over TRUSformer. Finally, the addition of the \emph{mixed} ensemble strategy in TRUSWorthy results in an additional increase of $1.6\%$ in AUROC when compared to a standard Ensemble of TRUSformers, and an increase of $2.5\%$ over TRUSformer, the previous SOTA in micro-ultrasound-based PCa detection. This improvement highlights the benefits of diversifying and increasing the data points presented to the ensemble members.

\begin{table}[t]
\caption{Performance of TRUSWorthy compared to prior work for PCa detection and uncertainty estimation. Metrics are averaged across folds, and the standard deviation is reported.}\label{tab:results_perf1}%
\begin{tabular}{@{}lcccccc@{}}
\toprule
\bf Method & AUROC$\uparrow$ & B.Acc.$\uparrow$ & Sens.$\uparrow$ & Spec.$\uparrow$ & ECErr$\downarrow$ & Brier$\downarrow$\\
\midrule
ResNet & $74.2 \pm 2.0$ & $68.1 \pm 1.7$ & $68.6 \pm 8.8$ & $67.6 \pm 6.9$ & $12.2 \pm 1.3$ & $21.7 \pm 1.5$ \\
Deep Ens.~\cite{lakshminarayanan2017simple} & $76.4 \pm 2.2$ & $67.7 \pm 3.5$ & $58.0 \pm 14$ & $77.5 \pm 8.7$ & $15.5 \pm 5.1$ & $21.7 \pm 1.6$ \\
SSL + ResNet \cite{wilson2023self} & $75.4 \pm 4.3$ & $66.2 \pm 5.1$ & $58.2 \pm 18$ & $74.2 \pm 9.0$ & $13.3 \pm 1.7$ & $21.7 \pm 1.5$ \\
SSL + Deep Ens. & $76.9 \pm 2.8$ & $68.0 \pm 3.9$ & $57.1 \pm 14$ & $\bf 78.9 \pm 7.7$ & $15.2 \pm 4.3$ & $20.5 \pm 1.3$ \\
\midrule
TRUSformer \cite{gilany2023trusformer} & $77.4 \pm 2.1$ & $69.5 \pm 5.4$ & $67.0 \pm 18$ & $72.0 \pm 12$ & $13.1 \pm 6.5$ & $21.6 \pm 7.0$ \\
TRUSformer Ens. & $78.3 \pm 2.1$  & $70.5 \pm 3.0$  & $66.1 \pm 11$ & $74.9 \pm 8.5$ & $5.59 \pm 2.7$ & $\bf 17.3 \pm 3.5$\\
TRUSWorthy &  $\bf 79.9 \pm 1.4$ & $\bf 71.5 \pm 0.7$ & $\bf 71.6 \pm 8.0$ & $71.3 \pm 6.8$ & $\bf 4.97 \pm 1.1$ & $17.4 \pm 1.7$ \\
\botrule
\end{tabular}
\end{table}

\begin{figure}[htbp] 
    \centering
    \subfloat[]{%
        \includegraphics[width=0.5\columnwidth]{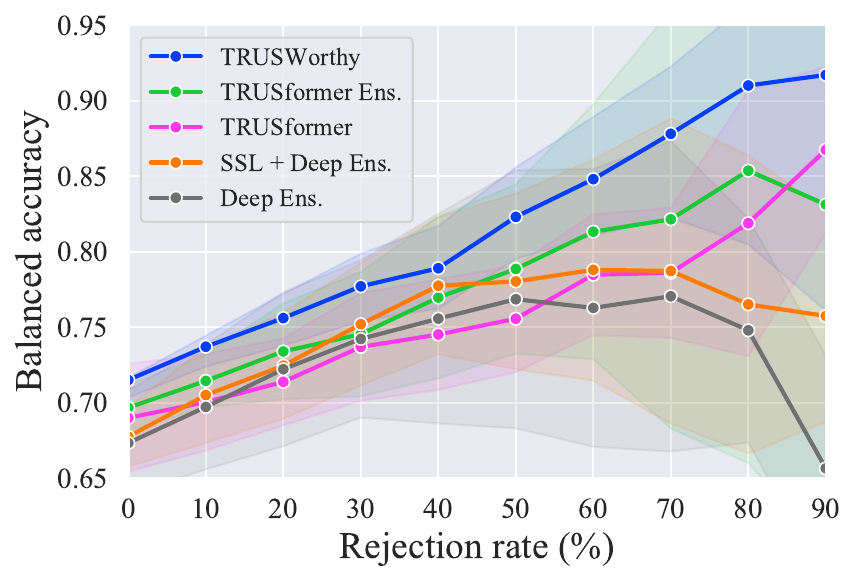}%
        \label{fig:rejection_analysis}%
        }%
    \subfloat[]{%
        \includegraphics[width=0.50\columnwidth]{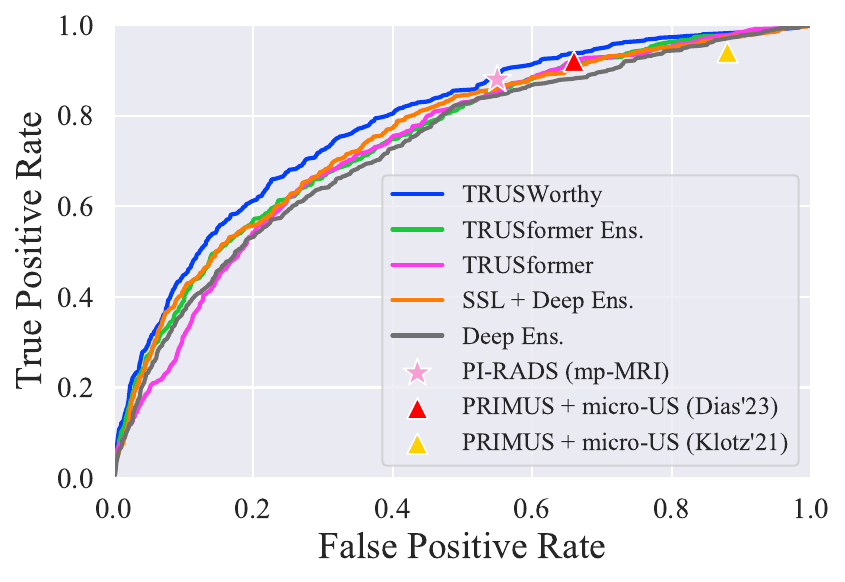}%
        \label{fig:roc_analysis}%
        }%
    \caption{\textbf{(a)} Accuracy-rejection plot showing the balanced accuracy of each method at different confidence thresholds. \textbf{(b)} ROC Curves for our method and other PCa detection baselines. True and False Positive Rates of clinical benchmarks are shown as points.}
\end{figure}

Overall, TRUSWorthy outperforms other methods in AUROC, balanced accuracy, sensitivity and ECErr. While it has nominally lower specificity ($\approx 7\%$) than other methods, ROC analysis (Figure \ref{fig:roc_analysis}) shows that its ROC curve lies above the other methods' at all true positive rates. This means that at higher detection thresholds, TRUSWorthy matches the specificity of these other methods and still achieves higher sensitivity.

When performing a leave-one-center-out evaluation approach, as shown in Table~\ref{tab:centerwise_metrics}, we can see TRUSWorthy's potential to generalize to new clinical settings when compared to other methods. We first observe that all evaluated methods perform relatively well on CRCEO and UVA, with AUROC results ranging between 78-84\%. TRUSWorthy slightly exceeds the performance of the strongest baseline method on 3 centers, and outperforms the single-model baselines significantly all centers. TRUSWorthy's performance on JH is significantly higher than any other method (+6\% compared to TRUSformer), leading us to conclude that this integrative method is an effective way to produce models that generalize to new clinical settings much better than any previous individual approach.

    \begin{table}[t]
    \centering
    \begin{tabular}{lccccc}
         \toprule
          & \multicolumn{5}{c}{\bf AUROC by Clin. Center} \\
          \bf Method & CRCEO & UVA & PMCC & PCC & JH \\
         \midrule
         SSL+ResNet  & 80.0 & 78.2 & 68.8 & 72.6 & 59.6 \\
         SSL+Deep Ens. & 78.6  & 79.2 & 69.6 & 75.3 & 59.9\\
         TRUSformer & 81.1 & 78.4 & 70.1 & 78.4 & 65.7 \\
         TRUSformer Ens. & 84.2 & 80.6 & \bf 75.0 & \bf 78.9 & 65.5 \\
         TRUSWorthy & \bf 84.4 & \bf 81.7 & 73.7 & \bf 78.9 & \bf 71.7 \\
         \bottomrule
    \end{tabular}
    \caption{Cancer detection performance of our 5 strongest models across different clinical centers, with leave-one-center-out validation.}
    \label{tab:centerwise_metrics}
    \end{table}

\noindent\textbf{\textit{Uncertainty Estimation: }} Table~\ref{tab:results_perf1} also reports the uncertainty estimation metrics of our models. Our main observation is the combined benefit of MIL and ensembles: in particular, TRUSformer-Ens. and TRUSWorthy both have markedly lower ECErr and Brier score than the other methods. TRUSWorthy outperforms TRUSformer-Ens. slightly in ECErr (-0.6\%), suggesting an additional benefit of having diversified ensemble members for uncertainty estimation.

Figure~\ref{fig:rejection_analysis} delineates the performance of the highest-performing models across varying uncertainty levels. The \( x \)-axis represents the \textit{rejection rate} \( r \), shown as a percentage. For a specified \( r \), we determine an uncertainty threshold that leads to the rejection of \( r \) percent of the samples with the highest uncertainty. We then compute the balanced accuracy for the remaining samples. The advantages of our method are evident here: as the rejection rate increases, there is a steady enhancement in accuracy. Notably, TRUSWorthy consistently outperforms other models at every uncertainty threshold. Especially striking is the leap from approximately \( 71\% \) accuracy at \( r=0\% \) to around \( 80\% \) at \( r=40\% \), which further escalates to surpass \( 90\% \) when \( r \) is \( 80\% \). This observation underscores the clinical significance of our results: a medical practitioner can fine-tune the uncertainty and detection thresholds to attain desired accuracy, sensitivity, and specificity levels. Whenever the model's predictions meet the clinician's confidence benchmarks, they can be employed; otherwise, conventional methods like systematic biopsy remain the fallback. The consistent performance of TRUSWorthy accentuates its apt naming; it shows its promise as a trustworthy tool for clinicians, ensuring reliable results in an uncertain clinical deployment environment.

\begin{figure}[t!] 
\centering
\includegraphics[width=0.89\columnwidth]{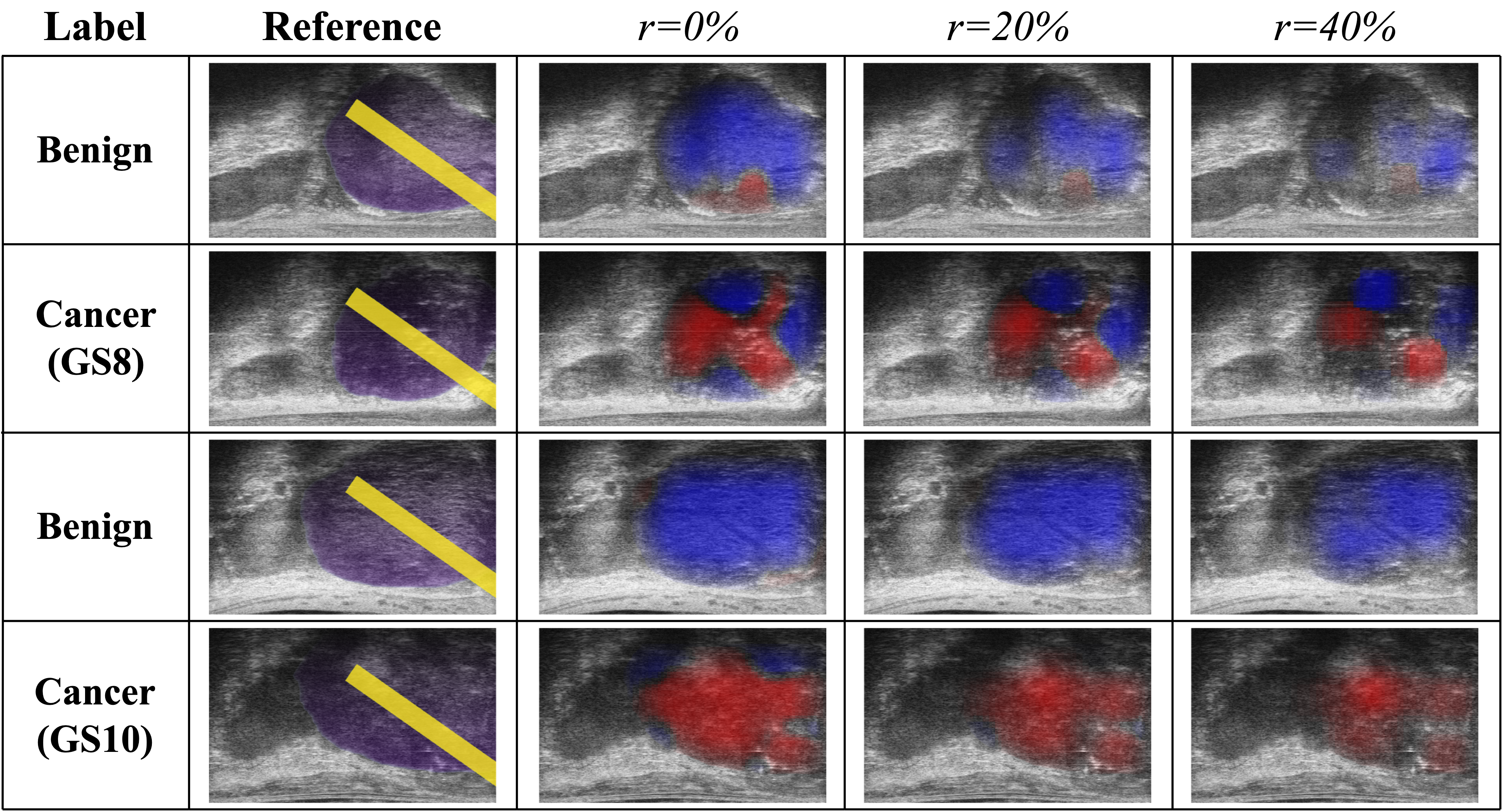}%
\caption{Visualizing TRUSWorthy's predictions at 3 rejection thresholds $(r=0,20,40\%)$. Cancer and benign predictions are highlighted in red and blue, respectively. The prostate and needle trace regions are shaded in purple and yellow, respectively.}
\label{fig:hmaps}%
\end{figure}

\noindent\textbf{\textit{Qualitative Analysis: }} Figure~\ref{fig:hmaps} shows the prediction heatmaps using TRUSWorthy overlaid on B-Mode prostate ultrasound images, with increasing rejection rates in columns from left to right. We showcase two benign examples and two examples of cancer with different Gleason scores. Overall, the output of the model is consistent with biopsy results. Rows 1 and 2 depict the utility of uncertainty thresholding: in Row 1, there are cancer predictions at low $r$-values which are rejected as ``uncertain" at higher $r$ values, and are most likely correctly rejected false positives. In Row 2, on the other hand, cancer predictions are confident, persisting at high $r$-values, and in this case matching the biopsy finding of a Gleason score 8 cancer. For Rows 3 and 4, the model predictions on the needle and surrounding prostate are confident and match the pathology results for the biopsy. 

\noindent\textbf{\textit{Comparison with clinical benchmarks: }} Given the impressive performance of TRUSWorthy compared to prior SOTA in AI-based PCa detection, we assess its potential for eventual clinical translation. Table~\ref{tab_clinical} compares TRUSWorthy's sensitivity, specificity and balanced accuracy to currently used visual detection methods PIRADS and PRIMUS~\cite{klotz2020can, klotz2021comparison,ahmed2017diagnostic} as reported in clinical studies. TRUSWorthy is competitive overall, with a higher specificity and balanced accuracy at the cost of a reduced sensitivity. This competitive performance suggests that TRUSWorthy may have a future role in clinical practice of prostate cancer biopsy by providing an objective and user-independent tool to complement existing visual detection methods. Because our present analysis does not control for differences in study populations and biopsy methodologies, a well-designed prospective validation study to rigorously compare PCa detection methods and establish the clinical efficacy of TRUSWorthy should be the focus of immediate future work.

\begin{table}[t]
\caption{Comparison of TRUSWorthy with clinical benchmarks for prostate cancer detection.}\label{tab_clinical}%
\begin{tabular}{@{}llllll@{}}
\toprule
\textbf{Method} & Patients & {Sens.} & {Spec.} & B. Acc. \\
\midrule
PI-RADS + mp-MRI~\cite{klotz2021comparison} & 576 & 88.0 & 45.0 &  66.5\\
PRIMUS + micro-US~\cite{klotz2021comparison} & 1040 & \bf{94.0} & 22.0 & 58.0\\
PRIMUS + micro-US~\cite{dias2023diagnostic} & 139 & 92.0 & 44.0 & 68.0\\
TRUSWorthy (ours) & 693 & $71.6$ & \bf{71.3} & \bf{71.5} \\
\botrule
\end{tabular}
\end{table}

\section{Limitations}
Our study excluded cores with low cancer involvement (below 40\% of the biopsy area). This decision is meant to ensure the model is geared toward clinically significant cancer detection from ultrasound, and to limit the number of false positives.
Moreover, Gleason Grade 1 (GG1) and Gleason Score 6 (GS6) cores, which represent clinically insignificant cancers, were not included in the finalized version of the dataset. However, it is important to acknowledge that GG1 and GS6 can potentially act as an intermediate class lying between benign and clinically significant cancers (Gleason Score $\geq$7). The elimination of this intermediate class may have simplified the task of differentiating cancer from benign, potentially inflating predictive performance and limiting the model’s clinical utility. In future studies, it may be worthwhile to consider using data with GG1 cores to provide a more nuanced challenge for the model and improve its generalizability in clinical settings. Furthermore, it is unknown whether the biopsy population consists of biopsy-naïve patients, those under active surveillance, or another group.

Finally, we utilized TRUS data exclusively, which may have resulted in undersampling PCa in the transition zone (TZ) and anterior gland. While our method is mainly intended for use in TRUS-guided biopsy, future work could improve the generalizability and robustness of the work by including data from transperineal ultrasound in order to study the detection of cancers in other zones of the prostate.

\section{Conclusion}\label{sec5}
We proposed TRUSWorthy, an uncertainty-aware framework for MIL classification of PCa in micro-ultrasound data. By simultaneously addressing weak labeling, data scarcity, class imbalance and heterogeneous data, TRUSWorthy achieves SOTA performance in deep learning PCa detection from micro-US. TRUSWorthy handles uncertainty by refraining from making uncertain predictions, making it a promising tool for robust and trustworthy PCa detection. 


\backmatter

\section*{Declarations}
This work was supported by the Natural Sciences and Engineering Research Council of Canada (NSERC), and the Canadian Institutes of Health Research (CIHR). Parvin Mousavi is supported by the CIFAR AI Chair and the Vector Institute. Brian Wodlinger is Vice President of Clinical and Engineering at Exact Imaging. None of the other authors have potential conflicts of interest to disclose. All patient data was used with informed consent and approval of institutional ethics boards.


\end{document}